\documentclass[english,aps,prl, twocolumn,superscriptaddress]{revtex4-1}
\usepackage[utf8]{inputenc}
\usepackage{amsmath}
\usepackage{amssymb}
\usepackage{graphicx}
\usepackage{hyperref}
\usepackage{siunitx}
\usepackage{placeins}
\usepackage{braket}
\usepackage{babel}
\usepackage{mathtools}

\newcommand{\mN}{m_{\rm N}}
\newcommand{\mS}{m_{\rm S}}
\newcommand{\kFS}{k_{\rm S}}
\newcommand{\kFN}{k_{\rm N}}
\newcommand{\kSz}{k_{{\rm S}z}}
\newcommand{\kNz}{k_{{\rm N}z}}
\newcommand{\vFS}{v_{\rm S}}
\newcommand{\vFN}{v_{\rm N}}
\newcommand{\vSz}{v_{{\rm S}z}}
\newcommand{\vNz}{v_{{\rm N}z}}
\renewcommand{\vr}{{\bf r}}
\newcommand{\sinc}{{\rm sinc\,}}

\begin{document}
\title{Proximity-induced gap in nanowires with a thin superconducting shell}
\author{Thomas Kiendl}
\affiliation{Dahlem Center for Complex Quantum Systems and Fachbereich Physik, Freie Universit\"at Berlin, 14195, Berlin, Germany}
\author{Felix von Oppen}
\affiliation{Dahlem Center for Complex Quantum Systems and Fachbereich Physik, Freie Universit\"at Berlin, 14195, Berlin, Germany}
\affiliation{Institute of Quantum Information and Matter, California Institute of Technology, Pasadena, California 91125, USA}
\author{Piet W. Brouwer}
\affiliation{Dahlem Center for Complex Quantum Systems and Fachbereich Physik, Freie Universit\"at Berlin, 14195, Berlin, Germany}

\begin{abstract}
Coupling a normal metal wire to a superconductor induces an excitation gap $\Delta_{\rm ind}$ in the normal metal. In the absence of disorder, the induced excitation gap is strongly suppressed by finite-size effects if the thickness $D_{\rm S}$ of the superconductor is much smaller than the thickness $D_{\rm N}$ of the normal metal and the superconducting coherence length $\xi$. We show that the presence of disorder, either in the bulk or at the exposed surface of the superconductor, significantly enhances the magnitude of $\Delta_{\rm ind}$, such that $\Delta_{\rm ind}$ approaches the superconducting gap $\Delta$ in the limit of strong disorder. We also discuss the shift of energy bands inside the normal-metal wire as a result of the coupling to the superconducting shell.
\end{abstract}

\maketitle

\section{Introduction} 

The creation of heterostructures is a powerful technique to combine effects that are otherwise hard to find in a single material. A combination that attracted considerable theoretical and experimental attention over the past decade is the simultaneous occurrence of superconducting pairing, spin-orbit coupling, and spin polarization in one-dimensional systems. The broad interest in these systems stems from the possibility that they may enter a phase of topological superconductivity. In a wire geometry, exponentially localized zero-energy Majorana bound states may appear at the boundary between topologically trivial and nontrivial regions,  with potential applications to topological quantum computation \cite{Kitaev2003, Nayak2008}. One setting which has been proposed for observing these effects relies on heterostructures consisting of a spin-orbit coupled nanowire and a superconductor  \cite{Lutchyn2010, Oreg2010}. Corresponding experiments have been performed in a variety of setups \cite{Mourik2012, Das2012, Churchill2013, Deng2016, Albrecht2016}. 

Recent experiments investigate nanowires proximitized by thin superconducting shells made of Al, with a thickness of the order of $\SI{100}{\nano\m}$ or less \cite{Krogstrup2015, Chang2015, Deng2016, Albrecht2016, Zhang2018}. Besides reducing the size of the experimental setup, these thin shells are advantageous as they reduce the magnetic flux through the superconductor for fields parallel to the wire, as they allow one to exploit charging energies for probing Majorana bound states \cite{Albrecht2016}, and as they can be epitaxially grown on top of the nanowire, which provides very clean interfaces between the two materials \cite{Krogstrup2015}. The latter is believed to be responsible for a hard proximity-induced gap at zero magnetic field, which has been observed in experiments \cite{Chang2015}. 

In view of the typical length scales of the system, these results may at first sight be rather surprising. Specifically, the coherence length of Al is in the $\mu \si{\m}$ range, much larger than the thickness of the superconducting coat. Thus, finite-size effects are expected to play a significant role. While early theoretical studies focused on nanowire-superconductor heterostructures for which finite-size effects can be neglected \cite{Sau2010, Duckheim2011, Zyuzin2013, Peng2015,  vanHeck2016}, more recent studies have considered the implications of a finite thickness of the superconductor. For a one-dimensional wire proximitized by thin two- or three dimensional superconducting coats, Reeg {\em et al.}\ suggested that finite-size effects can be detrimental to the induced gap \cite{Reeg2017, Reeg2018}.
Other works considered the effects of spatially-varying electrostatic potentials. Under suitable conditions, this may cause charge accumulation at the wire-superconductor interface and thus promote the proximity effect by pushing the wave function inside the nanowire closer to the interface \cite{Antipov2018, Mikkelsen2018, Woods2018}. 

In experiments, the interface between the epitaxially grown Al and the nanowire is expected to be relatively clean \cite{Krogstrup2015}, but the exposed Al surface might introduce a sizable amount of disorder or surface roughness. In the literature, disorder has been studied for wide superconductors coupled to nanowires, with disorder present in the wire  \cite{Akhmerov2011, Brouwer2011, Diez2012, Liu2012, Stanescu2012, Bagrets2012, Stanescu2011, Rainis2013}, the wire surface, \cite{Stanescu2011, Sau2012} at the end of the wire \cite{Pientka2012}, and inside the superconductor \cite{Stanescu2011, Cole2016}. The recent study by Reeg {\em et al.}\ investigated nanowires proximitized by  a thin, disordered superconducting layer, but found only a weak enhancement of the induced gap in the presence of moderate disorder strengths \cite{Reeg2018}. In addition, these authors find a large energy-shift of the nanowire bands due to coupling to the superconductor.

In this work, we investigate thin two- and three dimensional superconducting coats (S) coupled to a single-mode nanowire (N), with a cross section as shown in Fig. \ref{fig:setup}. Here, ``thin'' means that the thickness $D_{\rm S}$ of the superconducting coat is small compared to the superconducting coherence length. Our goal is to understand the consequences of the finite thickness, the dimensionality, and the disorder (both in the bulk and at the surface) of the superconductor. We go beyond previous works in the literature \cite{Reeg2018} by using a continuum model for the wire and the superconductor, so that --- within the limits imposed by a continuum description with quadratically dispersing bands --- the role of the device geometry can be assessed in our calculations.

The remainder of our work is structured as follows. In Sec.\ \ref{sec:basic_picture}, we give a qualitative discussion of the magnitude of the induced gap and the induced band shift of the nanowire bands. We introduce the continuum model used for the detailed calculations in Sec.~\ref{sec:model}. Section \ref{sec:clean} contains the analysis of the continuum model in the absence of disorder. In Sec.~\ref{sec:disorder_analytics} we include disorder in our discussion and derive analytical estimate for the proximity-induced gap $\Delta_{\textrm{ind}}$ by using a semiclassical ansatz. We compare with a numerical solution of the continuum model in Sec.~\ref{sec:scattering_approach}. Finally, we conclude in Sec.~\ref{sec:conclusion}.

\section{Qualitative discussion}\label{sec:basic_picture}

\begin{figure}
	\begin{center}
\includegraphics[width=0.8\columnwidth]{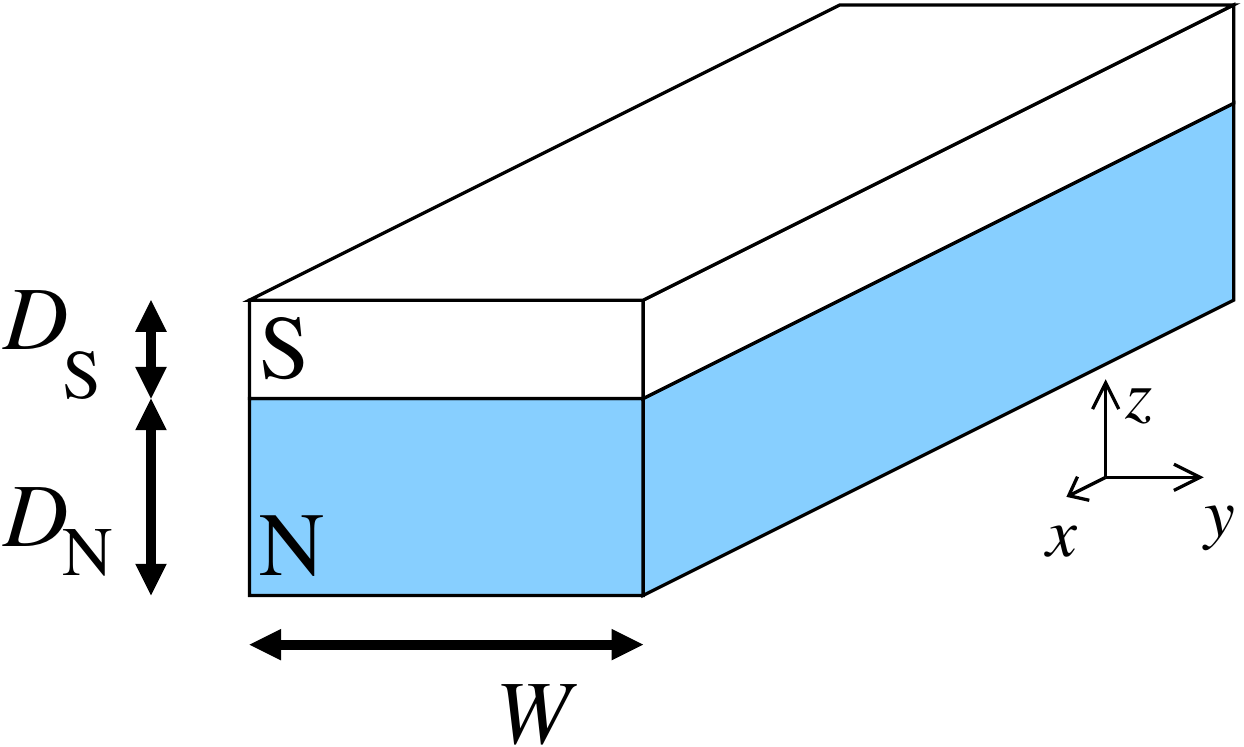}\bigskip

\includegraphics[width=\columnwidth]{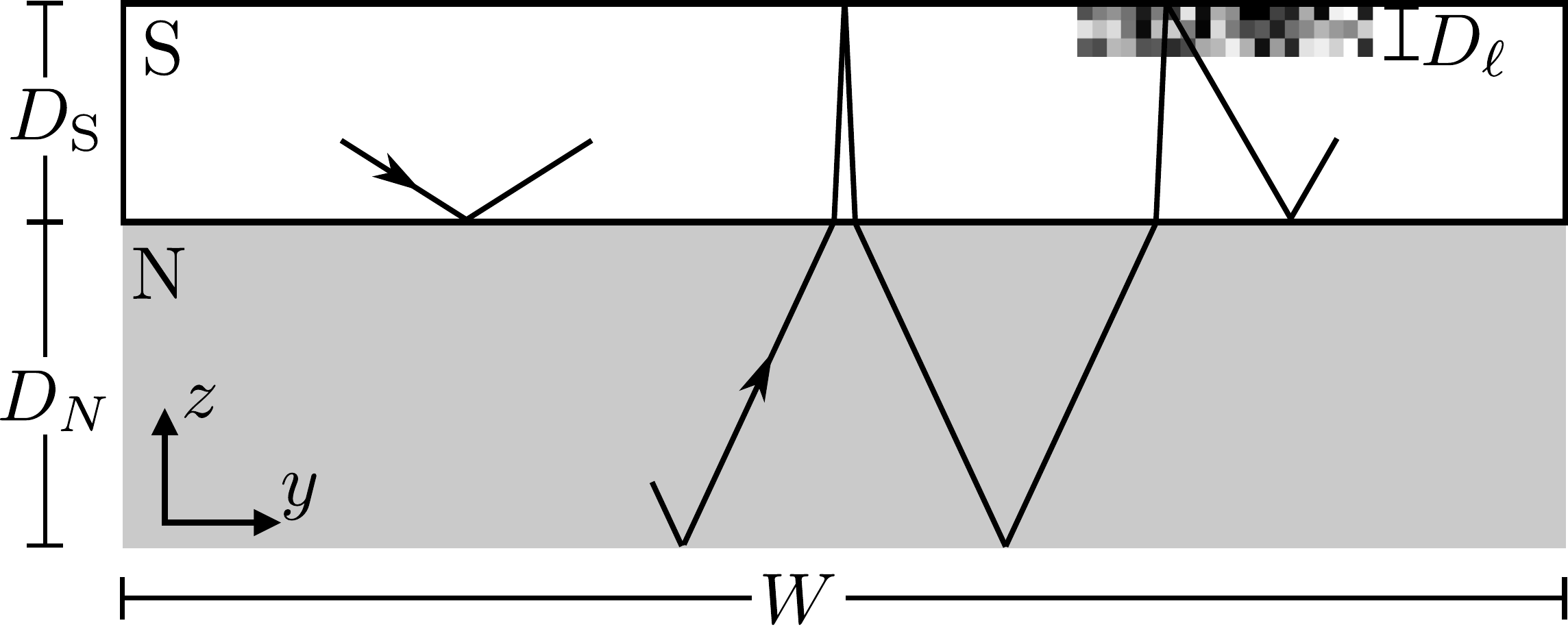}
\end{center}
\caption{\label{fig:setup}
A normal-metal wire (N) of thickness $D_{\rm N}$ coated by a thin superconductor (S) of thickness $D_{\rm S}$ (top). The sample width is $W$. The bottom panel shows a cross section of the devide along the $xz$ plane, together with the relevant semiclassical scattering processes. The left trajectory shows quasiparticles hitting the interface away from normal incidence. In such a case there is total internal reflection because of the large wavenumber mismatch between S and N. Transmission through the NS interface takes place only if the trajectory is close to normal incidence on the S side (center). Surface disorder, indicated in the top right, scatters modes that can enter N into modes that are totally reflected at the interface.}
\end{figure}

We model the semiconductor-superconductor heterostructure as a bilayer wire consisting of a normal metal (N) of thickness $D_{\rm N}$ and a superconducting layer of thickness $D_{\rm S}$. This setup is shown schematically in Fig.\ \ref{fig:setup} (top). We choose coordinate axes such that the $z$ axis is perpendicular to the NS interface and the $x$ axis points along the wire, see Fig.\ \ref{fig:setup}. The Fermi wavenumber $\kFS$ in the superconductor is much larger than the Fermi wavenumber $\kFN$ in the semiconductor, reflecting the vastly different electron densities in the two layers. At the same time, the Fermi velocities $\vFS$ and $\vFN$ are comparable, allowing (in principle) for the possibility of a strong coupling between the two layers, since the interface transparency depends on the ratio $\vFS/\vFN$. The thickness $D_{\rm S}$ of the superconducting layer is much smaller than the superconducting coherence length 
\begin{equation}
  \xi = \frac{\hbar \vFS}{\Delta},
\end{equation}
with $\Delta$ being the magnitude of the superconducting gap. We further assume that $D_{\rm S} \lesssim D_{\rm N}$, consistent with the typical experimental device geometry.

For a sufficiently small pairing potential $\Delta$, a description of the transverse modes of the NS bilayer can be obtained starting from the case of a ``metal-metal junction'' for which $\Delta=0$ inside S. Within a semiclassical picture and in the absence of disorder, the wavefunctions of such a metal-metal junction correspond to propagating electron or hole states, with quantized transverse momenta in the $y$ and $z$ directions. Superconductivity only plays a role at lengthscales $\gtrsim \xi$, at which electrons propagating in S are retroflected into holes and vice versa. The time required for this retroreflection process $\hbar/\Delta$ may be identified with the inverse superconducting gap. 
In a hybrid normal-metal--superconductor system the time required for reflection of electrons into holes and vice versa is longer than $\hbar/\Delta$, because the time spent in the normal region has to be added. Consequently, the induced gap $\Delta_{\rm ind}$ is reduced below $\Delta$.

We first  estimate the magnitude of $\Delta_{\rm ind}$ for the setup of Fig.~\ref{fig:setup} in the absence of disorder. In this case, the momenta in the $x$ and $y$ directions are preserved. For fixed $k_x$ and $k_y$ and in the limit of a small interface transparency modes occur at discrete energies only, corresponding to states localized almost entirely within S or N with $k_z$ quantized in steps of $\pi/D_{\rm S}$ or $\pi/D_{\rm N}$, respectively. Generically a mode propagating in N will couple off-resonantly to S. Hence, these modes have little overlap into S and the induced gap becomes small, which is in agreement with the findings of Ref.\ \cite{Reeg2017}. In this regime the magnitude of the induced gap may fall well below the induced gap in the limit of a normal-metal wire coupled to a bulk superconductor ($D_{\rm S} \to \infty$), for which the coupling between modes in N and S is described by Fermi's Golden Rule.

Next, we consider an interface transparency close to unity, which requires approximately matching Fermi velocities in S and N. Because of the large difference in electronic density the wavenumbers $\kFS \gg \kFN$ remain vastly different, however. As a result, quasiparticles transmitted into S from N will propagate almost perpendicularly to the interface, as shown schematically in the bottom panel of Fig.\ \ref{fig:setup}. Correspondingly, quasiparticles in S that approach the NS interface at normal incidence will be transmitted with large probability, whereas quasiparticles incident at generic angles are reflected. Hence, although the superconductor has a much larger density of states than the normal-metal wire --- as follows from the condition $\kFS \gg \kFN$ --- most of these states are effectively decoupled from N. For a mode in $N$ with a velocity $\vNz$ in the $z$ direction, the fraction of the time spent in S is $(D_{\rm S}/\vFS)/(D_{\rm S}/\vFS + D_{\rm N}/\vFN)$, which leads to 
\begin{equation}\label{eq:clean_gap_t1}
\Delta_{\rm ind} = \frac{\Delta}{1 + D_{\rm N}\vFS/D_{\rm S} \vNz}.
\end{equation}
Since the velocities are approximately matching and as $D_{\rm N}/D_{\rm S}$ is typically large, $\Delta_{\rm ind}$ is still small, compared to the bulk gap $\Delta$.

The induced gap can be significantly enhanced by the inclusion of disorder in S. For unit transparency, after an electron propagating in N enters S, disorder can scatter it out of the narrow range of angles normal to the NS interface for which a strong coupling at the NS interface exists. Once such scattering has occurred it is unlikely that the quasiparticle be scattered back into a range of angles for which it can return to the normal metal, as shown schematically in the bottom panel of Fig.~\ref{fig:setup}. In such a case an electron-like quasiparticle will be retroreflected as hole after a time $\hbar/\Delta$ (and vice versa). In this strong-disorder limit the total rate of scattering from an electron into a hole, and hence the induced gap, becomes
\begin{equation}\label{eq:dirty_gap_t1}
\Delta_{\rm ind} = \frac{\Delta}{1 + 2 D_{\rm N}\vFS/\xi \vNz},
\end{equation}
where the factor two appears because the time spent in N is $2 D_{\rm N}/\vNz$.
For current experiments that typically use aluminium as a superconductor \cite{Krogstrup2015, Chang2015, Deng2016, Albrecht2016, Zhang2018}, $\xi$ is much larger than $D_{\rm N}$. This gives an induced gap of order $\Delta$, which is in agreement with experimental observations. Although a similar reasoning appears in Ref.\ \cite{Reeg2018}, there the conclusion was that only a small increase of the induced gap is possible.

Finally, let us discuss the energy shift induced in the nanowire bands as a result of the coupling to the superconductor. For an isolated nanowire, the transverse modes are quantized, in the simplest case with a momentum $\hbar \pi/D_\textrm{N}$ perpendicular to the interface if the coupling to the superconductor is weak. The zero-point energy associated with quantization of $k_z$ (as well as quantization of $k_y$ --- although the latter is not affected by the coupling to the superconductor) raises the energies of states in N. Increasing the coupling to S effectively increases $D_{\rm N}$ and thus leads to a decrease of the energy offset from transverse confinement. The relative importance of this ``band shift'' depends on the interface transparency and on the thicknesses $D_{\rm S}$ and $D_{\rm N}$, as we discuss in detail in Appendix \ref{app:band_shift}. However, for typical experimental parameters it remains well below the initial finite-size shift associated with zero-point motion in the $z$ direction.

In the case of Ref.\ \cite{Reeg2018}, the nanowire is modeled as being effectively two-dimensional without extension in the $z$ direction ({\em i.e.}, effectively by setting $D_{\rm N} = 0$). The corresponding energy shift from size quantization in the $z$ direction is absent in such a model, which explains why the authors of Ref.\ \onlinecite{Reeg2018} could have arrived at the conclusion that the band shift from coupling to S is appreciable in their model. In recent experiments on nanowires coated by Al, one typically has $D_\textrm{N} \gg D_\textrm{S}$ and hence the induced band-shift is expected to be small \cite{Krogstrup2015, Chang2015, Deng2016, Albrecht2016, Zhang2018}. However, we also note that in these systems multiple transverse bands might cross the Fermi level and the electrostatic potential is expected to be nontrivial \cite{Antipov2018, Mikkelsen2018, Woods2018}, which makes a quantitative comparison with experiments difficult.

\section{Continuum model}\label{sec:model}

We now describe our quantitative calculations using a continuum model for a normal-metal wire with a superconducting shell. The system under consideration is shown in Fig.\ \ref{fig:setup}. As described in the previous Section, we consider a normal-metal (N) wire of thickness $D_{\rm N}$ coupled to a thin superconducting layer (S) of thickness $D_{\rm S}$. We choose coordinate axes such that the $x$ and $z$ directions are along the wire and perpendicular to the NS interface, respectively. The interface between the two materials is located at $z=0$ and both materials are restricted to $0<y<W$. The $2 \times 2$ Bogoliubov-de Gennes Hamiltonian reads
\begin{equation}\label{eq:hamiltonian_full}
  \hat{\mathcal{H}} = 
  \begin{pmatrix}
  \hat{H}_0 & \theta(z) \Delta \\
  \theta(z) \Delta & - \hat{H}_0^*
  \end{pmatrix},
\end{equation}
for a spinor wavefunction $\psi = (u,v)^{\rm T}$ consisting of particle and hole wavefunctions of opposite spin. We choose the gauge such that the superconducting order parameter $\Delta$ is real and positive. The Heaviside step function $\theta(z) = 1$ ($0$) for $z > 0$ ($z < 0$). The normal-state Hamiltonian $\hat{H}_0$ reads
\begin{equation}
  \hat{H}_0 = \xi_{\bf p}(z) + V_{\rm conf}(y, z) + U(\vr).
  \label{eq:H0}
\end{equation}
We consider the parabolic dispersion
\begin{equation}\label{eq:normal_dispersion}
	\xi_{\bf p}(z) = 
  \sum_{\alpha = x,y,z} p_{\alpha} \frac{1}{2 m_{\alpha}(z)} p_{\alpha}
  + V_0(z),
\end{equation}
where we take the mass tensor $m_{\alpha}$ to be isotropic in the superconductor, 
\begin{equation}
  m_{\alpha}(z) = m_{\rm S},\ \ \alpha = x,y,z,\ \ \mbox{for $z > 0$},
\end{equation}
whereas we allow for an anisotropic mass in the normal metal,
\begin{equation}
  m_{x}(z) = m_{{\rm N}x},\ \ m_{y}(z) = m_{z}(z) = m_{\rm N} \ \
  \mbox{for $z < 0$}.
\end{equation}
The potential $V_0(z)$ is a band offset, which we parameterize in terms of Fermi wavenumbers $\kFS$ and $\kFN$ for the superconductor and the normal metal, respectively,
\begin{align}
  V_0(z) =&\, - \frac{\hbar^2 \kFS^2}{2 m_{\rm S}}\ \ \mbox{for $z > 0$}, \\
  V_0(z) =&\, - \frac{\hbar^2 \kFN^2}{2 m_{\rm N}}\ \ \mbox{for $z < 0$}.
\end{align}
The anisotropic mass for the N region is introduced for technical reasons in order to simplify our numerical calculations, see the discussion in Sec.\ \ref{sec:scattering_approach}. It has no consequences for the qualitative conclusions. The confining potential $V_{\rm conf}(y, z)$ models the sample boundary, $V_{\rm conf}(y, z) = 0$ for $-D_{\rm N} < z < D_{\rm S}$ and $0 < y < W$, and $V_{\rm conf}(y, z) = \infty$ otherwise. We assume disorder to be present at the exposed top boundary of the superconductor at $z = D_{\rm S}$, with an extension over a region of width $D_{\ell}$ into the superconductor. We model the corresponding potential $U(\vr)$ as Gaussian white noise with zero mean and with correlation function
\begin{equation} \label{eq:disorder_correlator}
  \left\langle U(\mathbf{r})U(\mathbf{r}')\right\rangle =
  \frac{\hbar v_{\rm S}}{2 \pi \nu_0 \ell}
  \delta
  \left(\mathbf{r}-\mathbf{r}'\right),
\end{equation}
with support for $0 \leq D_{\rm S} - D_{\ell} \leq z \leq D_{\rm S}$ only. 
Here, $v_{\rm S} = \hbar \kFS / \mS$ and the densities of states per spin direction in two and three dimensions read $\nu_0 =  \kFS/2 \pi \hbar v_{\rm S}$ and $\nu_0 = \kFS^2/2\pi^2\hbar v_{\rm S}$, respectively. The parameter $\ell$ corresponds to the mean free path in the disorder region if $\kFS \ell \gtrsim \pi$. Strong surface scattering corresponds to the regime $\ell \ll D_{\ell}$.

\section{Induced gap without disorder}
\label{sec:clean}

\subsection{Transverse modes in the absence of superconductivity}\label{sec:metal_metal}

As a starting point for our calculations we first consider the case $\Delta = 0$ corresponding to a junction of two normal metals. We calculate the propagating modes in the absence of the disorder potential $U(\vr)$. These will form the basis of our subsequent analysis.

We write the wavefunction, normalized to unit flux along the $x$ direction, as
\begin{align} \label{eq:metal_metal_wave_function}
	\psi_{\nu} (\vr, \varepsilon) &= 
	\frac{
		e^{ i s \tau k_x(\tau \varepsilon) x}\sin \frac{\pi n_y}{W} y
	}{
		\sqrt{ W \hbar v_x/2}
	}
	\varphi_{\tau, n_z}(z, \varepsilon)
	,
\end{align}
where the multi-index $\nu = (s, \tau, n_y, n_z)$ labels the direction of propagation $s=\pm$, the electron/hole sector $\tau$, and the positive integer quantum numbers $n_y$ and $n_z$ counting the quantized momenta in the transverse directions. Further $k_x$ is the longitudinal momentum, which is real and positive, and $v_x = \hbar |d\varepsilon/dk_x|$. We write $\tau = e(h)$ when it appears as an index and $\tau = 1(-1)$ otherwise. The transverse mode functions $\varphi$ read
\begin{align}\label{eq:psi_perp_def_1}
	\varphi_{e, n_z}(z, \varepsilon) &= 
	\frac{
		c_{\rm e} e^{i \kNz(\varepsilon) z} + c_{\rm e}' e^{-i \kNz(\varepsilon) z}	
	}{
		\sqrt{\vNz(\varepsilon) \mathcal{N}_{e, n_z}}
	}, \\ \label{eq:psi_perp_def_2}
	\varphi_{h, n_z}(z, \varepsilon) &=
	\frac{
		c_{\rm h} e^{-i \kNz^*(-\varepsilon) z} + c_{\rm h}' e^{i \kNz^*(-\varepsilon) z}	
	}{
		\sqrt{\vNz(-\varepsilon) \mathcal{N}_{h, n_z}}
	}
\end{align}
for $z < 0$ and
\begin{align}\label{eq:psi_perp_def_3}
	\psi_{\perp \tau, n_z}(z) &= 
	\frac{
		d_{\rm \tau} e^{-i \tau \kSz z} + d_{\rm \tau}' e^{i \tau \kSz z}	
	}{
		\sqrt{\vSz \mathcal{N}_{\tau, n_z}}
	}
\end{align}
for $z > 0$. Here, $\mathcal{N}_{\tau, n_z}$ are normalization constants such that $\int d z |\varphi_{\tau, n_z}(z)|^2 =1 $ and
\begin{align}
	\kNz(\varepsilon) &= \sqrt{\kFN^2 - k_y^2 - \frac{\mN}{m_{\textrm{N}x}}k_x^2 + 2 \mN \varepsilon/\hbar}
	\label{eq:kNz}
	, \\
	\kSz(\varepsilon) &= \sqrt{\kFS^2 - k_y^2 - k_x^2 + 2\mS \varepsilon/\hbar}
	\label{eq:kSz}	
	, \\
	\vSz &= \hbar \kSz/m_{\rm S}
	, \\
	\vNz &= \hbar \kNz/m_{\rm N}
	,
\end{align}
where we have dropped the multi-index $\nu$ for the wavenumbers and velocities. We recall that the large electron density in S implies that $\kFS \gg \kFN$, so that $\kSz$ is real, whereas $\kNz$ may be complex.

Upon requiring continuity of $\psi$ and its spatial derivative at the NS interface at $z=0$, one finds that the coefficients $c_{\tau}$ and $c'_{\tau}$ satisfy the conditions \cite{Kupferschmidt2009}
\begin{align}\label{eq:match_interface}
	\begin{pmatrix}
		c_{e }' \\ d_{e }' \\ c_{h }' \\ d_{h }'
	\end{pmatrix}
	=
	\begin{pmatrix}
		r & t' & 0 & 0 \\
		t & r' & 0 & 0 \\
		0 & 0 & r^* & (t')^* \\
		0 & 0 & t^* & (r')^*
	\end{pmatrix}
	\begin{pmatrix}
		c_{e} \\ d_{e} \\ c_{h} \\ d_{h}		
	\end{pmatrix}
\end{align}
with the interface transmission and reflection amplitudes
\begin{align}
	\label{eq:tUp}
 	t &= \frac{2 \sqrt{\vSz \vNz}}{\vSz + \vNz},  
 	\\
	\label{eq:tpUp}
 	t' &= \frac{2 \sqrt{\vNz \vSz}}{\vSz + \vNz},  
  	\\ \label{eq:rUp}
 	r &= -1 + t \sqrt{\vNz/\vSz},
 	\\ \label{eq:rpUp}
  	r' &= -1 + t' \sqrt{\vSz/\vNz}.
\end{align}
For our later analysis it is useful to define the transmission amplitude at normal incidence
\begin{equation}
t_{\perp} = \frac{2 \sqrt{\vFS \vFN}}{\vFS + \vFN},
\end{equation}
where $\vFS = \hbar \kFS/\mS$ and $\vFN = \hbar \kFN/\mN$. The boundary conditions at $z = -D_{\rm N}$ and $z=D_{\rm S}$ yield the additional conditions
\begin{align}\label{eq:match_wire_wall1}
	c_{e}' &= - c_{e} e^{-2 i \kNz(\varepsilon) D_{\rm N}} ,\\
	c_{h}' &= - c_{h} e^{2 i \kNz^*(-\varepsilon) D_{\rm N}}, \\
	d_{\tau}' &= - d_{\tau} e^{- 2i \tau  \kSz(\tau \varepsilon) D_{\rm S}}.
	\label{eq:match_wire_wall2}
\end{align}

Equations \eqref{eq:match_interface} and \eqref{eq:match_wire_wall1}--\eqref{eq:match_wire_wall2} fully determine the wavefunction, and yield the transcendental equation
\begin{align} \label{eq:kz_quantization}
	0 &= \vNz \cot \kNz(\tau \varepsilon) D_{\rm N} + \vSz \cot \kSz(\tau \varepsilon) D_{\rm S}.
\end{align}
Solution of Eq.~(\ref{eq:kz_quantization}) yields the quantized values for $\kSz(\tau \varepsilon)$, which we label using the integer index $n_z$. We set $c_{\tau} = 1$, which fixes the remaining $c$- and $d$-coefficients via Eqs.\ \eqref{eq:match_interface} and \eqref{eq:match_wire_wall1}--\eqref{eq:match_wire_wall2} and leads to
\begin{align}\label{eq:norm}
	\mathcal{N}_{\tau, n_z} =&\, 
	\frac{2 D_{\rm S}}{\vSz} |d_{\tau}|^2
	\left(
		1 - \frac{\sin (2 \kSz D_{\rm S})}{2 \kSz D_{\rm S}}
	\right)
	\\&\, \nonumber \mbox{}
	+
	\frac{e^{4 D_{\rm N} {\rm Im} \kNz} - 1}{2 {\rm Im} \kNz |\vNz|}
	\\&\, \nonumber \mbox{}
	- e^{2 D_{\rm N} {\rm Im} \kNz} \frac{\sin (2 D_{\rm N} {\rm Re} \kNz) }{ |\vNz| {\rm Re} \kNz}. 
\end{align}
The dependence on $\tau \varepsilon$ was dropped on the right hand side of this equation for the sake of compactness.

\subsection{Excitation gap}\label{sec:clean_system}

To calculate the excitation gap, we start start by expressing the Hamiltonian \eqref{eq:hamiltonian_full} in the basis of the propagating modes \eqref{eq:metal_metal_wave_function}. We assume that $\Delta$ is the smallest energy scale in the problem. This implies that $\xi/D_{\rm S}$, $\xi/W \gg 1$, such that the transverse modes are well described by the transverse components for $\Delta = 0$, see Eq.~\eqref{eq:metal_metal_wave_function}. Moreover, in this limit we can safely assume that the transverse modes are non-degenerate, so that the effect of the superconducting pairing $\Delta$ can be treated for each transverse subband separately.

Calculating the mode spectrum using degenerate first-order perturbation theory, we find that the transverse mode with quantum numbers $n_y$ and $n_z$ has dispersion
\begin{equation}
  \varepsilon = \pm \sqrt{\varepsilon_{n_y,n_z}(k_x)^2 + |\Delta_{n_y,n_z}|^2},
\end{equation}
where
\begin{align}
	\varepsilon_{n_y,n_z}(k_x) =&\, 
  \frac{\kSz^2 + (n_y \pi/W)^2 + k_x^2 - \kFS^2}{2 \mS},
	\\
	\Delta_{n_y,n_z} =&\, \Delta  \int_0^{D_{\rm S}} dz
  \varphi_{e,n_z}(z,0)^* \varphi_{h,n_z}(z,0), \label{eq:gap_alpha_after_AA}
\end{align}
where the energy argument of the transverse mode functions $\varphi_{\tau,n_z}(z,\varepsilon)$ has been set to zero. We conclude that the overall gap of the system is
\begin{equation}\label{eq:gap_ind_definition}
	\Delta_{\rm ind} = \underset{n_y,n_z}{\rm min} |\Delta_{n_y,n_z}|
\end{equation}

Upon evaluating Eq.~\eqref{eq:gap_alpha_after_AA}, we obtain
\begin{equation}\label{eq:clean_ind_gap}
\Delta_{n_y,n_z} =  \frac{2 D_{\rm S}\Delta}{\vSz \mathcal{N}_{\alpha}} |d_{n_y,n_z}|^2
	\left(
		1 - \frac{\sin 2 \kSz D_{\rm S}}{2 \kSz D_{\rm S}}
	\right),
\end{equation}
where
\begin{align}\label{eq:d4}
	|d_{n_y,n_z}|^2 =&\left|
	\frac{
		\vNz \cos \kNz D_{\rm N}- i \vSz \sin \kNz D_{\rm N}
	}{
		\vSz \cos \kSz D_{\rm S} - i \vNz \sin \kSz D_{\rm S}
	}\right| 
  \nonumber \\ =&\, \mbox{} \times
	e^{-{\rm Im} \kNz D_{\rm N}},
\end{align}
Due to the exponential factor in Eq.~\eqref{eq:d4}, which also appears in the normalization factor $\mathcal{N}_{\tau,n_z}$, modes evanescent in the normal region will have $\Delta_{n_y,n_z} \approx \Delta$, with corrections that are exponentially suppressed in ${\rm Im} \kNz D_{\rm N}$. (Note that we use the convention ${\rm Im} \kNz < 0$.) In the following discussion of limiting cases, we focus on modes with real $\kNz$. Furthermore, we consider the regime of a single mode inside the normal region below the Fermi level, $1 \lesssim\kFN D_{\rm N}/\pi < 2$ and $1 \lesssim\kFN W/\pi < 2$, set $m_{{\rm N} x} = \mN$, and take the limits $\kFS \gg \kFN$, $\mS \gg \mN$ and $\kFS D_{\rm S}\gg 1$, which are the relevant parameter regimes of current experiments on semiconductor-superconductor hybrids.

First, we discuss the case of unit transparency, which corresponds to $\vNz = \vSz$. In the limit $\kFS \gg \kFN$, we have $\kSz \approx \kFS$ and since $\vNz$ lies between $0$ and $\vFN$, we require $\vFN > \vFS$ for this to occur. Furthermore, the parameters have to be tuned in order to fulfill Eq.~\eqref{eq:kz_quantization}, which yields
\begin{align}
	\kFS D_{\rm S} + \vFS \mN D_{\rm N} &= n \pi,
\end{align} 
with $n$ a positive integer. In order to obtain $\Delta_{\rm ind}$ from Eq.~\eqref{eq:gap_ind_definition} we argue that since we consider the regime of a single mode in the wire, and since the remaining modes in the superconductor have an evanescent overlap with the wire, the mode with unit transparency will have the minimum magnitude of the mode-specific gap $|\Delta_{n_y,n_z}|$. After evaluating Eq.~\eqref{eq:clean_ind_gap}, we obtain
\begin{equation}\label{eq:peak_gap_t1}
	\Delta_{\rm ind} = \Delta \left[1 + \frac{ D_{\rm N} }{D_{\rm S} }\left(1 - \frac{\sin 2 \kNz D_{\rm N}}{2 \kNz D_{\rm N}} \right) \right]^{-1}.
\end{equation}

Equation~\eqref{eq:peak_gap_t1} agrees with Eq.~\eqref{eq:clean_gap_t1} for matching velocities, up to the interference term in parentheses. In the derivation of Eq.~\eqref{eq:clean_gap_t1}, we assumed a classical propagation of electrons and holes, which explains the absence of this interference term. Furthermore, we expect $\kNz D_N \sim \pi$, in which case the interference term is a numerically small correction.

\begin{figure}
	\begin{center}
	\includegraphics[width=\columnwidth]{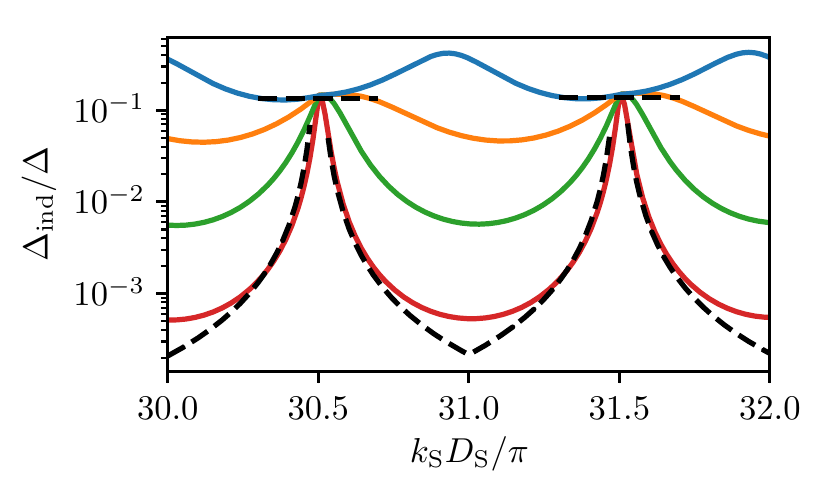}
	\end{center}
	\caption{\label{fig:clean_gap_ind}
	Induced gap in the absence of disorder. We choose a velocity mismatch $\vFN/\vFS = 0.1$, $0.33$, $1$, $3.3$ for the lowest (red) to the highest lying solid line (blue), respectively. The dashed lines show the prediction at the peak \eqref{eq:peak_gap_t1}, and the tails \eqref{eq:tail_gap}. The remaining parameters are $\kFN D_{\rm N} = 1.8 \pi$, $\kFS/\kFN = 100$. 
}
\end{figure}

Next, we consider the induced gap in the small-transparency limit $\vFS/\vFN \gg 1$ and for $\kFS^2/\kFN^2 \gg \kFS D_{\rm S}$. In order to solve Eq.~\eqref{eq:kz_quantization}, we expand around the solution for zero transparency and at the Fermi level,
\begin{align}
	\kSz &= \left (n_{\rm S} + \frac{1}{2}\right) \frac{\pi}{D_{\rm S}} + \delta q,\\
	 \kNz &= \frac{\pi}{D_{\rm N}} + \delta k,
\end{align}
with
\begin{align}
	n_{\rm S} &= \lfloor k_{\rm S} D_{\rm S}/\pi \rfloor,
	\\	
	\delta q &= k_{\rm S} - \left(n_{\rm S} + \frac{1}{2}\right) \frac{\pi}{D_{\rm S}},
\end{align}
where $\lfloor \cdot \rfloor$ denotes the floor function. Here, we neglected the contribution of $\delta k$ to $\kSz$, which is justified in the limit $\kFN^2/\kFS^2 \gg \kFS D_{\rm S}$.
Expanding Eq.~\eqref{eq:kz_quantization} to lowest order in $\delta q$ and $\delta k$ yields,
\begin{equation}\label{eq:quantization_expanded}
	\frac{\pi}{\mN D_{\rm N}^2 \delta k} + \frac{1}{\mN D_{\rm N}}
	= 
	\vFS \delta q,
\end{equation}
which has the solution 
\begin{equation}\label{eq:dk_wire_high_mismatch}
	\delta k = \frac{\pi}{\mN D_{\rm N}^2 \vFS \delta q D_{\rm S} - D_{\rm N}}.
\end{equation}
For $1 \gtrsim \delta q D_{\rm S} \gg \vFN/\vFS$, both $\delta q$ and $\delta k$ are sufficiently small to justify the
lowest order expansion around the zero transparency case. Finally, we argue that for $\delta_q D_{\rm S} \gtrsim {\vFN D_{\rm S}}/{\vFS D_{\rm N}}$ the remaining propagating modes have evanescent overlap into the wire and can be neglected. Thus, the solution \eqref{eq:dk_wire_high_mismatch} is expected to be the transverse mode with the minimal gap value and from Eq.~\eqref{eq:gap_ind_definition} we obtain
\begin{equation}\label{eq:tail_gap}
	\Delta_{\rm ind} = \frac{\Delta \pi^2}{D_{\rm N}^3 \vFS^2 \mN^2 \delta q^2 D_{\rm S}}.
\end{equation}

Figure \ref{fig:clean_gap_ind} shows numerical results obtained by solving Eqs. \eqref{eq:kz_quantization} and \eqref{eq:gap_ind_definition}. The resulting effective gap as a function of $\kFS D_{\rm S}$ is an approximately $\pi$-periodic function. The induced gap is sharply peaked near half-integer values of $\kFS D_{\rm S}/\pi$ if $\vFN \ll \vFS$. In this regime we find good agreement with Eq.~\eqref{eq:tail_gap}. The peak value at half-integer values of $\kFS D_{\rm S}/\pi$ is well approximated by Eq.~\eqref{eq:peak_gap_t1}. Upon increasing $\vFN$ the peaks a half-integer $\kFS D_{\rm S}/\pi$ become broader and asymmetric. The peak structure inverts when $\vFN \approx \vFS$, resulting in a gap function with maxima close to integer values of $\kFS D_{\rm S}/\pi$ when $\vFN \gg \vFS$.

\section{Disordered hybrid NS wires} \label{sec:disorder_analytics}

\subsection{Qualitative considerations}

As discussed in the introduction, disorder scatters between modes with support in the semiconductor wire and modes predominantly localized in the superconductor. Consequently, the time spent in N is reduced, which increases the induced gap $\Delta_{\rm ind}$.

To estimate the induced gap in the presence of disorder and for a transparent interface (matching Fermi velocities in N and S), we consider the case of a single propagating mode in N and assume that the remaining transverse modes are entirely localized in S. (These modes decay exponentially on the N side of the NS interface.) The mode relevant for the induced gap is the one propagating in N, and we estimate the induced gap as 
\begin{equation}\label{eq:induced_gap_semiclassical}
\Delta_{\rm ind} = \frac{1}{\Delta^{-1} + t_{\rm N}/\hbar},
\end{equation}
where $t_{\rm N}$ is the time an electron spends in N between two Andreev reflections.

In order to estimate $t_{\rm N}$, we consider a reference point along an electron trajectory in the normal metal and calculate $t_{\rm N}$ as sum of the mean times in N until the next Andreev reflection and the mean time since the previous Andreev reflection. Without Andreev reflection, such a trajectory alternatingly makes round trips through the normal metal (from $z=0$ to $z=-D_{\rm N}$ and back) and through the superconductor (from $z=0$ to $z = D_{\rm S}$ and back). The probability that the round trip through the superconductor results in Andreev reflection is
\begin{equation}
	P_{\rm A} = 1-e^{ - 4 D_{\rm S}/\xi -2 D_{\rm S}/\ell_{\rm eff}}.
\end{equation}
Here $e^{-4 D_{\rm S}/\xi-2 D_{\rm S}/\ell_{\rm eff}}$ is the probability that the electron is neither retroreflected into a hole during the round trip, nor disorder-scattered into a different mode. As discussed previously, if disorder scattering occurs, the electron is scattered into a state that is not coupled to the normal metal with probability close to unity, and Andreev reflection takes place with unit probability. In the case of a uniform disorder strength throughout the superconductor we have $\ell_{\rm eff} = \ell$. For surface disorder (modeled by disorder confined to a strip of width $D_{\ell}$), we set 
\begin{equation}\label{eq:ell_eff_surface}
  \ell_{\rm eff} = \frac{D_{\rm S}}{D_{\ell}} \max(\ell, a_{\rm sat} \pi/\kFS),
\end{equation}
where $a_{\rm sat}$ is a numerical constant of order one and is determined numerically in Appendix~\ref{sec:appendix:surf_disorder_saturation}. Equation (\ref{eq:ell_eff_surface}) describes the saturation of scattering in the limit of strong disorder at the superconductor's exposed surface.

Each time the electron is \emph{not} retroreflected as a hole, an additional time ${2 D_{\rm N}}/{\vNz}$ has to be spent inside N. For the time $t_{\rm N}$ between Andreev reflections we then find
\begin{equation}
	t_{\rm N} = \frac{2 D_{\rm N}}{\vNz}
  \left[1 + 2 (1 - P_{\rm A}) + 2 (1 - P_{\rm A})^2 + \ldots \right].
\end{equation}
Evaluating Eq.~\eqref{eq:induced_gap_semiclassical}, we obtain
\begin{equation}\label{eq:gap__semi_classical}
	\Delta_{\rm ind} = \Delta\left( 1 + \frac{2\vFS D_{\rm N} (2 - P_{\rm A})}{\vNz \xi P_{\rm A}}\right)^{-1}.
\end{equation}
In the no-disorder limit $\ell_{\textrm{eff}} \gg \xi \gg D_{\textrm{S}}$, Eq.~\eqref{eq:gap__semi_classical} reduces to Eqs. \eqref{eq:clean_gap_t1}, and in the limit $\xi \gg D_{\rm S} \gg \ell_{\rm eff}$, it turns into Eq.~\eqref{eq:dirty_gap_t1}. For intermediate disorder strengths, $\xi \gg \ell_{\textrm{eff}} \gg D_{\rm S}$, we obtain
\begin{equation}
	\Delta_{\textrm{ind}} = \Delta \left(1 + \frac{2 \vFS D_{\rm N} \ell_{\textrm{eff}}}{\vNz D_{\rm S} \xi} \right)^{-1},
\end{equation}
which grows monotonically upon increasing the disorder strength $\ell_{\rm eff}^{-1}$. 

In our derivation of Eq.~\eqref{eq:gap__semi_classical}, we assumed that once an electron scatters from disorder, it does not enter again into N. However, an electron might scatter from disorder and enter N one or multiple times. If $n_{\textrm{S}}$ modes are present in the superconductor, we expect these processes to become relevant for $\xi \gtrsim n_{\textrm{S}} \ell_{\textrm{eff}}$ only. This is also the scale at which Anderson localization is expected to occur, and hence we cannot access this regime with our semiclassical approach. We note however, that the width $W$ can be increased to increase the number of modes and push the onset of this regime to larger coherence lengths.

\subsection{Numerical results}\label{sec:scattering_approach}

We compare the estimate \eqref{eq:gap__semi_classical} to a numerical calculation of the density of states. As before, we consider the geometry of Fig. \ref{fig:setup} and the continuum Hamiltonian \eqref{eq:hamiltonian_full}. We consider a hybrid NS wire of length $L$, numerically determine the scattering matrix $S(\varepsilon)$ (see App.\ \ref{sec:appendix:numerical_method} or Refs.\ \onlinecite{Brouwer2003,Bardarson2007,Sbierski2014} for details), and calculate the density of states using the relation
\begin{equation}\label{eq:def_dos}
	\rho(L,\varepsilon) = 
  \frac{1}{2 \pi i} {\rm Tr} S(L, \varepsilon)^\dagger \frac{d S(L, \varepsilon)}{d \varepsilon}.
\end{equation}
The calculation of the scattering matrix requires that source and drain leads are added to the system. The leads are described by the same Hamiltonian as the hybrid NS wire, but without the disorder potential $U(\vr)$ and the pairing potential $\Delta$. We infer the size of the induced gap by noting that if the system is gapped and $\varepsilon$ lies above the gap, $\rho(L,\varepsilon)$ is proportional to $L$. In contrast, for $\varepsilon$ inside the gap, $\rho(L,\varepsilon)$ converges to an $L$-independent residual density of states for $L \to \infty$ since the lead modes partially extend into the wire. 

The numerical analysis of the original problem is complicated by the fact that the transverse mode functions $\varphi_{\tau, n_z}(z)$ are in general non-orthogonal if the masses in the N and S regions are different. (Note that the full wavefunctions in \eqref{eq:metal_metal_wave_function} still form an orthonormal set.) In order to circumvent this problem, we take the mass in the N region to be anisotropic, with $m_{{\rm N}x} = \mS$. With this choice, the transverse mode functions are orthogonal as a function of $z$, at fixed $n_y$ and $\tau$. For the case of a single mode in N this change does not qualitatively alter the results of our analysis: Setting $m_{{\rm N}x}$ merely gives a constant energy offset for the single propagating mode in N, whereas the other modes are evanescent in N and are hardly affected by this substitution.

\begin{figure}
	\begin{center}
\includegraphics[width=\columnwidth]{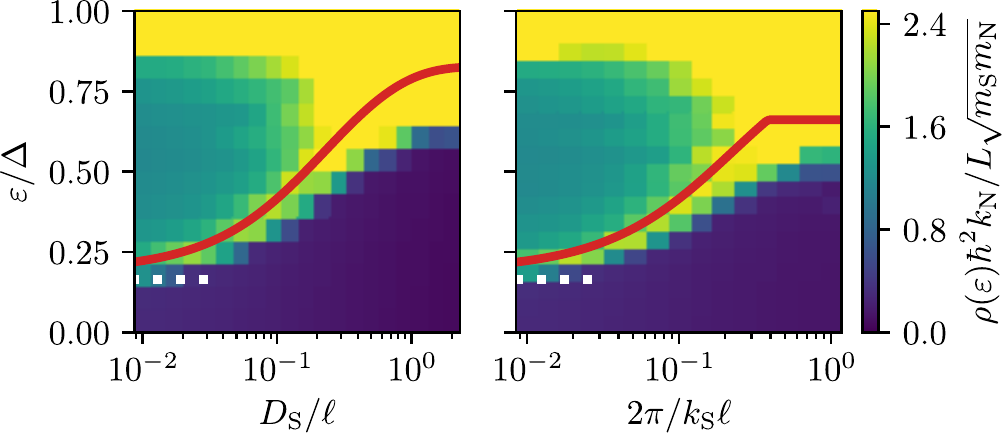}
\end{center}
\caption{\label{fig:2d_eps_vs_ell}
Density of states as a function of energy and disorder strength for a single-mode semiconducting wire coupled to a two-dimensional superconductor. We choose $\xi/D_{\rm S} = 40$, $\vFN/\vFS = 1.5$ and disorder located over the full width of the superconductor (left, $D_{\ell} = D_{\rm S}$) and the top surface (right, $D_{\ell} = 2 \pi /\kFS$). The white dots show Eq.~\eqref{eq:peak_gap_t1} and the red line shows Eq.~\eqref{eq:gap__semi_classical} with $a_{\rm sat} = 5.2$. The remaining parameters are $k_{\rm S} D_{\rm S} = 20.4 \pi$, $\mS/\mN = 100$, $L/\xi = 8$ and $k_{\rm N} D_{\rm N} = 1.2\pi$. The density of states is averaged over 5 disorder realizations. Values exceeding the color scale are mapped to the maximum value of the colorbar.
}
\end{figure}

For a two dimensional system, extended in the $xz$ plane, the density of states obtained from the numerical calculation is shown in Fig.~\ref{fig:2d_eps_vs_ell}. For all disorder strengths a gap is visible, indicated by the dark region. For small disorder strengths a Van Hove singularity clearly indicates the edge of the induced gap, and at $\varepsilon \sim \Delta$ the high density of states reflects the bulk gap. Furthermore, for very weak disorder, the induced gap converges to the value predicted by Eq.~\eqref{eq:peak_gap_t1} (white dots). The induced gap starts to increase once the effective mean free path exceeds the coherence length $\xi$ and saturates when $\ell$ becomes comparable to the thickness $D_{\rm S}$ of the superconducting layer, in good agreement with Eq.~\eqref{eq:gap__semi_classical} (red line). 

At the strongest disorder values $\Delta_{\rm ind}$ remains slightly below the limiting value of Eq.~\eqref{eq:clean_gap_t1}. We attribute this smaller value as well as the decrease of the bulk gap in the same regime to the onset of Anderson localization in our numerical simulations. The localization length $\xi_{\rm loc}$ is approximately given by $n_{\rm S} \ell_{\rm eff}$, where $n_{\rm S}$ is the number of transverse modes. For the results shown in Fig.~\ref{fig:2d_eps_vs_ell} the number of transverse modes $n_{\rm S} \sim 20$, so that we indeed expect localization effects to play a role at the largest disorder strengths considered in the figure. We note that Anderson localization is not expected to play a significant role in numerical simulations of three dimensional systems, where the number $n_{\rm S}$ of transverse modes is typically much larger.

\begin{figure}
	\begin{center}
\includegraphics[width=.7\columnwidth]{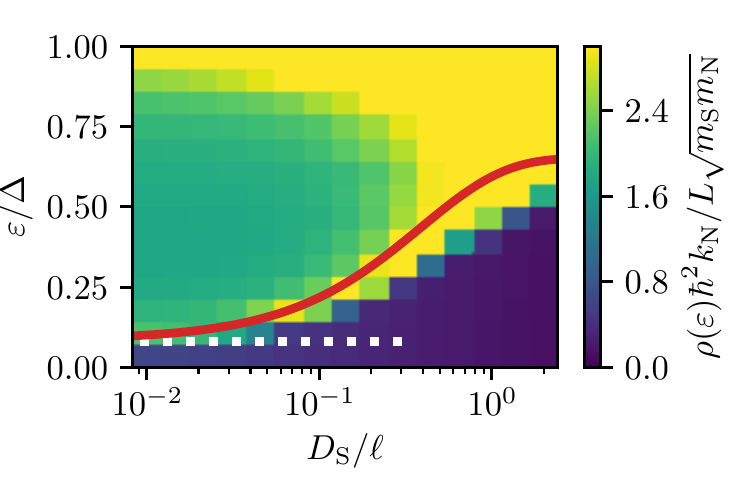}
\end{center}
\caption{\label{fig:3d_eps_vs_ell}
Density of states as a function of energy and disorder strength for a single-mode semiconductor wire coupled to a three-dimensional superconductor. We set $\kFS D_{\rm S}/\pi = 2.4$, $\kFS W = 18.7 \pi$, $\kFN W = 1.4 \pi$, and $k_{\rm N} D_{\rm N} = 1.2\pi$ such that only a single mode has a non-evanescent overlap into N.
We choose $\xi/D_{\rm S} = 40$, $\vFN/\vFS = 1.5$ and disorder located over the full width of the superconductor ($D_{\ell} = D_{\rm S}$).  The white dots show Eq.~\eqref{eq:peak_gap_t1} and the red line shows Eq.~\eqref{eq:gap__semi_classical}. The remaining parameters are $\mS/\mN = 20$ and $L/\xi = 22$. The density of states is averaged over 10 disorder realizations. Values exceeding the color scale are mapped to the maximum value of the colorbar.
}
\end{figure}

Next, we consider a three-dimensional superconductor with a larger number of modes along the $y$ direction in S, of which only the mode with $n_y=1$ has non-evanescent overlap into the normal region. The results are shown in Fig.~\ref{fig:3d_eps_vs_ell}. Again, the addition of disorder leads to a pronounced increase of the induced gap $\Delta_{\rm ind}$. We attribute this enhancement to the large number of modes along the $y$ direction $(n_y \leq 18)$, since only a small number of modes is present along the $z$ direction ($\kFS D_{\textrm{S}}/\pi = 2.4$). Furthermore, the results show good agreement with Eqs. \eqref{eq:peak_gap_t1} (red line) and \eqref{eq:gap__semi_classical} (white dots).

\section{Conclusion}\label{sec:conclusion}

In this work, we have investigated a normal-metal (N) wire coated by a thin two- or three dimensional superconductor (S), with disorder in the bulk or at the bare surface of the superconductor. Here ``thin'' means that the thickness $D_{\rm S}$ of the superconductor is much smaller than the superconductor coherence length $\xi$.

The coupling to the superconductor induces a gap $\Delta_{\rm ind}$ in the excitation spectrum of the normal metal. In the absence of disorder and for small interface transparencies, we find that this induced gap is much smaller than the induced gap for the case of a normal-metal wire coupled to a half-infinite superconducting shell, up to resonances that occur periodically when a momentum-preserving coupling between a level in the superconductor and the wire mode at the Fermi level is possible. Although the induced gap increases upon approaching a transparent interface, which requires matching Fermi-level velocities in N and S and removing the interface barrier, the induced gap is still smaller than the gap in the case of a half-infinite superconductor, the suppression factor being proportional to the ratio $D_{\rm N}/D_{\rm S}$ of the thickness of the N and S layers, which is typically large in experiments \cite{Krogstrup2015, Chang2015, Deng2016, Albrecht2016, Zhang2018}. 

Our results in the absence of disorder are in qualitative agreement with Ref.\ \cite{Reeg2017}, which studies a one-dimensional wire coupled to a thin superconductor. In Ref.\ \onlinecite{Reeg2017}, the coupling to the superconductor is described by a tunneling energy scale $\gamma$; our present approach features a continuum model, for which the coupling is described by the interface transmission probability $|t_{\perp}|^2$ at perpendicular incidence. For weak coupling the two quantities are related by $\gamma \sim |t_\perp|^2 \vFN / D_{\textrm{N}}$, and we find that our prediction for the suppression of the induced gap in Eq.~\eqref{eq:tail_gap} agree with those of Ref.\ \onlinecite{Reeg2017} up to a prefactor of order unity [see Eq.~(17) in Ref. \cite{Reeg2017}]. For unit transparency, our results predict a suppression of $\Delta_{\rm ind}$ by a factor $D_{\rm N}/D_{\rm S}$ as compared to the case of a half-infinite superconducting shell. No such suppression was found in Ref.\ \cite{Reeg2017}, in which $\Delta_{\rm ind}$ approaches the bulk gap $\Delta$ in the limit of strong coupling. 

A large band shift induced by the superconductor has been reported for a closely related tight-binding model in Ref.\ \onlinecite{Reeg2018}. In Appendix \ref{app:band_shift} we relate parameters in our continuum model to the lattice model of Ref.\ \onlinecite{Reeg2018}. While we quantitatively reproduce the band shift observed in Ref.\ \onlinecite{Reeg2018}, we also find that the shift quickly decays upon increasing the thickness $D_{\rm N}$ of the normal-metal wire and that it remains below the zero-point band offset $\varepsilon_0 = \hbar^2 \pi^2/2 m_{\rm N} D_{\rm N}^2$ at all times. While we expect that this band shift, which is attributed to a change of the zero-point confinement energy of electrons in the normal metal wire, has a relatively small effect if $D_{\rm N} \gtrsim D_{\rm S}$, other effects, not taken into account in our simple model analysis, such as interaction-induced band bending \cite{Antipov2018, Mikkelsen2018, Woods2018}, might be significant for a realistic modeling of current experiments \cite{Krogstrup2015, Chang2015, Deng2016, Albrecht2016, Zhang2018}.  

In the presence of disorder and for approximately matching Fermi velocities in N and in S, we find that disorder in the bulk or at the surface of the superconductor can significantly enhance the induced gap. We find that this enhancement sets in, when the effective mean free path $\ell_\textrm{eff}$ in the superconductor becomes smaller than the coherence length $\xi$. For the typical case when $\xi$ is large compared to the thicknesses $D_{\textrm{N}}$ and $D_{\textrm{S}}$ of N and S, we find an induced gap comparable to $\Delta$ for $\ell_{\textrm{eff}}/\xi \lesssim D_{\textrm{S}}/D_{\textrm{N}}$. We note that the condition $\ell_{\rm eff} \ll D_{\rm S}$ is met in most present experiments \cite{Chang2015, Deng2016, Albrecht2016, Zhang2018}, due to bulk disorder or to disorder at the exposed surface of the superconductor. In this regime, our results indicate that there is good reason to expect an induced gap of order $\Delta$, in spite of $D_{\rm S}$ being much smaller than $\xi$. This conclusion goes beyond the findings of Ref.\ \onlinecite{Reeg2018}, which only finds a weak enhancement of the induced gap in the presence of moderate disorder.

Whereas the quality of the NS interface has proven to be highly beneficial to inducing a topological superconducting state in semiconductor/superconductor hybrids, our results suggest that it is not desirable that the superconducting shells are ideal throughout. In particular, strong scattering at the exposed surface of the superconductor is essential to ensure a large magnitude of the induced superconducting gap.

\begin{acknowledgements}
\begin{acknowledgements}
We thank Bj\"orn Sbierski and Christian Kl\"ockner for discussions. Financial support was provided by the Deutsche Forschungsgemeinschaft DFG in the framework of project C03 of the Collaborative Research Center Transregio 183 entangled states of matter and by a QuantERA grant. One of us (FvO) thanks the Aspen Center for Physics (supported by National Science Foundation grant PHY-1607611) and the IQIM, an NSF physics frontier center funded in part by the Moore Foundation for hospitality while some of this work was performed. 
\end{acknowledgements}
\end{acknowledgements}

\appendix
\renewcommand\thefigure{\thesection.\arabic{figure}}   
\setcounter{figure}{0}

\section{Saturation for strong surface disorder}\label{sec:appendix:surf_disorder_saturation}

In this section, we discuss the crossover from weak ($\kFS \ell \ll 1$) to strong surface disorder ($\kFS \ell \lesssim 1$). We expect that scattering from surface disorder saturates when $\ell$ becomes small compared to the Fermi wavelength.

We verify this numerically, by considering a two dimensional metal slab of thickness $D_{\rm S}$ and length $L$, that is connected to two leads. Disorder is present only at one of the surfaces, for $D_{\rm S} - D_{\ell} < z < D_{\rm S}$, extending a distance $D_{\ell}$ into the system. The metal is described by the Hamiltonian \eqref{eq:hamiltonian_full} for $0 < z < D_{\rm S}$, with $\Delta = 0$ and without the normal region N.

\begin{figure}
\includegraphics[width=\columnwidth]{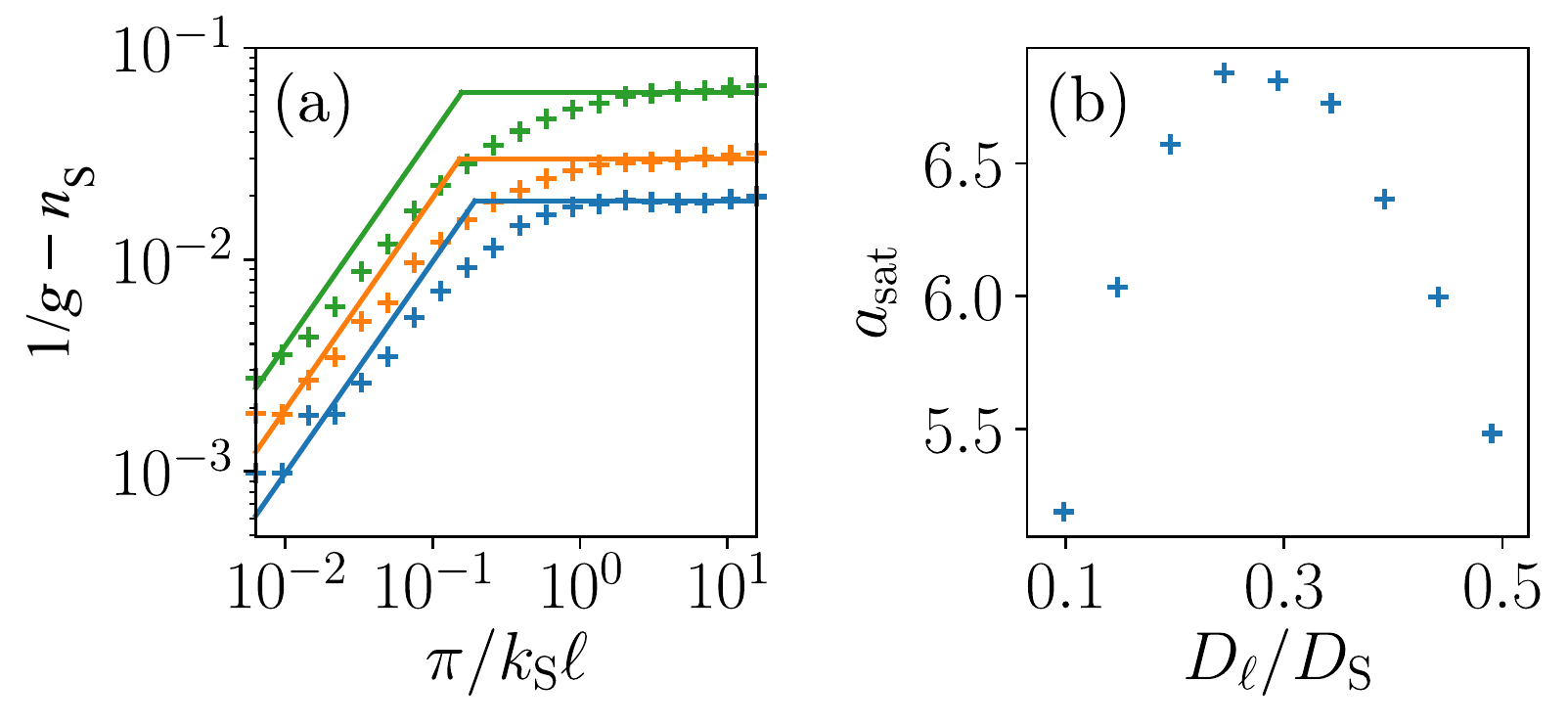}
\caption{\label{fig:G_saturation_dirty_surface} Saturation of effective scattering rate for different thicknesses of the disorder region. We choose a normal metal wire of width $\kFS D_{\rm S} = 20.4 \pi$ and $L = D_{\rm S}$. In (a), $\kFS D_{\ell}/\pi = 2$ (bottom, blue markers), $4$, and $8$ (top, green markers). The horizontal lines show the average of $1/g - n_{\rm S}$ for $\pi/\kFS \ell \geq 1$, the inclined lines show Eq. \eqref{eq:OhmsLaw}. The conductance is averaged over 40 disorder realizations. In (b), the coefficient $a_{\rm sat}$ is shown, as defined in Eq. \eqref{eq:ell_eff_surface}.}
\end{figure}

We define the effective mean free path $\ell_{\rm eff}$ by comparing the dimensionless conductance $g$ of this system with the conductance of a diffusive metal with uniform disorder of mean free path $\ell_{\rm eff}$,
\begin{equation}
  g_{\rm diffusive} = \frac{n_{\rm S}}{1 + L/\ell_{\rm eff}},
  \label{eq:OhmsLaw}
\end{equation}
where $n_{\rm S} = k_{\rm S} D_{\rm S}/\pi$ is the number of propagating modes at the Fermi level. Equation (\ref{eq:OhmsLaw}) is compared to a numerical calculation of the dimensionless conductance $g$ using the scattering approach, see App.\ \ref{sec:appendix:numerical_method} for details. The result of this comparison is shown in Fig.\ \ref{fig:G_saturation_dirty_surface}a for a fixed length $L$, which is small enough to avoid localization corrections. For sufficiently weak disorder ({\em i.e.,} for large $\ell$), Eq.\ (\ref{eq:OhmsLaw}) is in good agreement with the numerically obtained conductance with $\ell_{\rm eff} = \ell$, while for $\pi/\kFS \ell \gtrsim 1$ the conductance saturates. (We recall that the bulk mean free path is not a fit parameter: It is determined by the disorder correlation function (\ref{eq:disorder_correlator}).) The horizontal lines show the ensemble average of the asymptotic value of $g^{-1} - n_{\rm S}$ for $\pi/\kFS \ell \geq 1$, from which the coefficient $a_{\rm sat}$ in Eq. \eqref{eq:ell_eff_surface} can be determined. Figure \ref{fig:G_saturation_dirty_surface}b shows that the values of $a_{\rm sat} \approx 6$, with only a weak dependence on the ratio $D_{\rm ell}/D_{\rm S}$.

\begin{figure}
\includegraphics[width=\columnwidth]{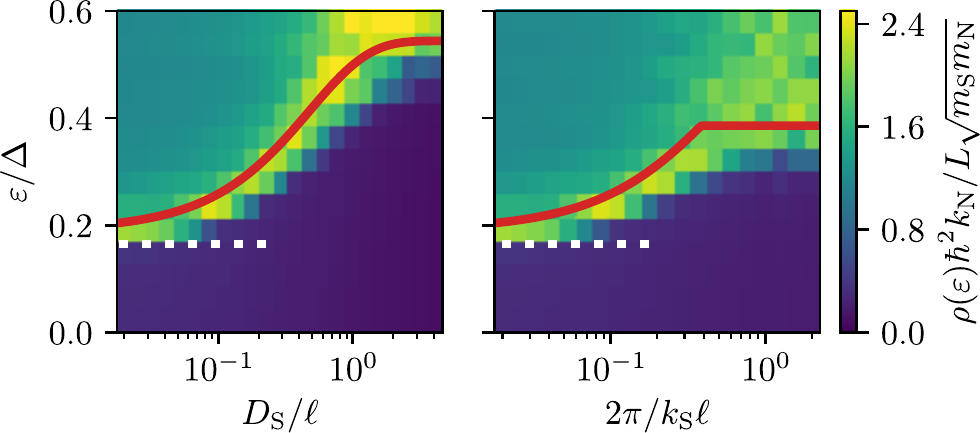}
\caption{\label{fig:2d_eps_vs_ell__small_xi}
Density of states as a function of energy and disorder strength for a 2d superconductor extended in the $x-z$ plane. We choose $\xi/D_{\rm S} = 10$, $\vFN/\vFS = 1.5$ and disorder located over the full width of the superconductor (left, $D_{\ell} = D_{\rm S}$) and the top surface (right, $D_{\ell} = 2 \pi /\kFS$). The white dots show Eq. \eqref{eq:peak_gap_t1} and the red line shows Eq. \eqref{eq:gap__semi_classical} with $a_{\rm sat} = 5.2$. The remaining parameters are $k_{\rm S} D_{\rm S} = 20.4 \pi$, $\mS/\mN = 100$, $L/\xi = 8$ and $k_{\rm N} D_{\rm N} = 1.2\pi$. The density of states is averaged over 8 disorder realizations. Values exceeding the color scale are mapped to the maximum value of the colorbar.
}
\end{figure}

The saturation of the effective scattering rate with increasing disorder strength for surface disorder scattering is mirrored in the saturation of the induced gap $\Delta_{\rm ind}$ upon increasing the disorder strength in the surface layer, as could be seen in the right panel of Figs.\ \ref{fig:2d_eps_vs_ell}. Figure \ref{fig:2d_eps_vs_ell__small_xi} shows a similar parameter configuration as in Fig. \ref{fig:2d_eps_vs_ell}, but with a smaller coherence length that significantly weakens the impact of Anderson localization at small $\ell$. In the case of surface disorder (right), Fig.\ \ref{fig:2d_eps_vs_ell__small_xi} shows a clear saturation for small $\ell$ which agrees well with Eq. \eqref{eq:gap__semi_classical}. In the case of bulk disorder (left), this saturation is absent and the induced gap reaches higher values.

\section{Relation to Ref. \cite{Reeg2018} and band shift} \label{app:band_shift}

Reeg {\em et al.} \cite{Reeg2018} considered a very similar system as the one we study in this article, but with a lattice model instead of a continuum description. In this appendix, we quantitatively compare our results to the lattice mdoel of Ref.\ \cite{Reeg2018} for the case that disorder is absent. 

Reference \cite{Reeg2018} describes the semiconducting nanowire as a lattice with a ($z$-direction) thickness of one site. The superconductor is modeled as a lattice with a finite thickness of multiple sites. As a result, in Ref.\ \onlinecite{Reeg2018} the thickness $D_{\rm N}$ of the normal wire does not enter as a separate parameter. On the other hand, in Ref.\ \onlinecite{Reeg2018} the coupling strength across the NS interface can be set by adjusting the hopping amplitude between the N and S regions.

One of the main findings of Ref.\ \cite{Reeg2018} is that the coupling to the superconductor can induce a large energy-shift in the nanowire bands. Below, we establish a quantative comparison between our continuum model and the lattice model of Reeg {\em et al.} We confirm that the band shift is also present in our model. However, since the thickness $D_{\rm N}$ of the normal metal explicitly enters into our model, we can study the $D_{\rm N}$-dependence of the band shift. Perhaps not surprisingly, we find that the band shft strongly decreases with increasing $D_{\rm N}$ and that at all times it remains below the zero-point band shift $\varepsilon_0 = \hbar^2 \pi^2/2 m_{\rm N} D_{\rm N}^2$ due to confinement in the $z$ direction, which was not considered in the lattice calculation of Ref.\ \onlinecite{Reeg2018}.

In order to quantitatively compare our model to Ref. \cite{Reeg2018} it is necessary that we extend the model studied in the main text to include an interface potential barrier and spin-orbit coupling. Hereto we add the terms
\begin{equation}
  \delta \hat H_0 = \hbar w \delta(z) + \sigma \alpha p_x \theta(-z)
  \label{eq:dH}
\end{equation}
to the normal-state Hamiltonian (\ref{eq:H0}). The first term introduces an interface potential, which allows us to tune the interface transparency independently of the Fermi velocities $\vFS$ and $\vFN$. The second term describes spin-orbit coupling restricted to the normal metal, with a strength $\alpha$ that couples the electron momentum to the electron spin $\sigma = \pm$. Following Ref.\ \onlinecite{Reeg2018} we neglect superconductivity in the discussion of the band shift. Furthermore, we assume an isotropic mass in the nanowire, $m_{\textrm{N}x} = m_\textrm{N}$. Since we are interested in the regime of a single band inside N that crosses the Fermi level, we may neglect any $y$ dependence, consider the bands with the lowest wavenumber $k_y$ only, and absorb any contributions from this lowest value of $k_y$ into the definitions of $\kFS$ and $\kFN$.

Upon inclusion of the additional terms (\ref{eq:dH}), the quantization condition \eqref{eq:kz_quantization} changes to 
\begin{equation}\label{eq:quantization_with_w}
	0 = 2 w + \vSz \cot \kSz D_{\textrm{S}} + \vNz \cot \kNz D_{\textrm{N}},
\end{equation}
where $\vSz = \hbar \kSz/m_{\rm S}$, $\vNz = \hbar \kNz/m_{\rm N}$, and
\begin{align}
	\kSz &= \sqrt{k_{\textrm{S}}^2 - k_x^2 + 2 m_{\textrm{S}} \varepsilon/\hbar^2}
	,\\
	\kNz &= \sqrt{k_{\textrm{N}}^2 - k_x^2 + 2 m_\textrm{N} (\varepsilon - \sigma \alpha \hbar k_x)/\hbar^2}
	.
\end{align}

To fit parameters of our continuum description to the model parameters of Ref.\ \onlinecite{Reeg2018} we match the dispersion of both models. The dispersion for the superconductor is readily matched, by equating the mass $\mS$, thickness $D_{\textrm{S}}$ 
\footnote{In our model, we adjust the thickness by one lattice site, $D_S = d + a$, where $d$ is the thickness of S and $a$ is the lattice spacing in Ref. \cite{Reeg2018}. This is motivated by considering the quantization of $\kSz$ in the absence of a normal region. In this case, the wave-number is quantized to multiples of $\pi/(d+a)$ in the tight-binding model and to multiples of $\pi/ D_\textrm{S}$ in the continuum model.}
and the velocity at the Fermi level $\vFS = \hbar \kFS/\mS$ in both models. To match parameters for the semiconductor wire we note that in our model, the dispersion of the semiconductor nanowire is controlled by the five parameters $\mN$, $D_N$, $\alpha$, $w$ and the band bottom $\hbar^2\kFN^2/2 \mN$. We choose the former three by equating $\alpha$ in both models, by matching the band curvature via tuning the mass $\mN$, and by setting $D_\textrm{N} = a$,  where $a$ is the lattice spacing in the tight-binding model of Ref.\ \cite{Reeg2018}. This leaves two parameters, $w$ and $\kFN$, which we determine by fitting making sure that the the bottom of the normal-wire band and the normal-wire band's avoided crossing with one of the bands of the superconductor are the same in both models. 

\begin{figure}
\includegraphics[width=\columnwidth]{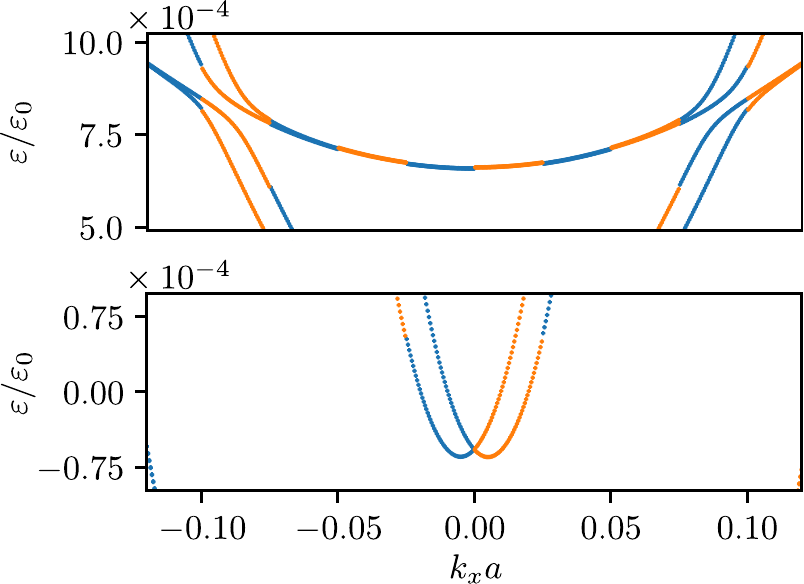}
\caption{\label{fig:app:dispersion_fitted_to_reeg} Fit of the dispersion used in this work (orange) to the dispersion in Ref. \cite{Reeg2018} Fig. 2 (b) (blue). The dispersions of the two models are shown in alternating intervals for better visibility. The orange segments show the solutions of Eq. \eqref{eq:quantization_with_w}. The parameters $w$ and $k_N$ are tuned such that the bottom of the N band at $k_x a \approx 0$ (lower panel) and the avoid crossing at $k_x a \approx \pm 0.1$ match, where $a$ is the lattice spacing in the model of Ref.\ \onlinecite{Reeg2018}. The parameters for our model are $D_{\rm N} = a$, $D_\textrm{S}/D_\textrm{N} = 43$, $\mS/\mN = 5$, $\kFS D_\textrm{S}/\pi = 4.35$, $\kFN D_\textrm{N}/\pi = 0.99201$, $w\mN/ \kFN = 19.88$ and $\alpha \mS / \kFS = 0.08$. The parameters for the tight-binding model are those of Fig. 2b of Ref.\ \cite{Reeg2018}. Figure 2 of Ref.\ \onlinecite{Reeg2018} shows $\varepsilon$ in units of $\Delta$ with the choice $\Delta = 10^{-3} \hbar^2 \kFS^2/2 m_{\rm S}$. We have chosen to show the dispersions in unit of the zero-point energy $\varepsilon_0 = \hbar^2 \pi^2/2 m_{\rm N} D_{\rm N}^2$ because $\Delta$ has been set to zero in our calculation of the dispersion, see the discussion in the text. We only show solutions with the lowest momentum $k_y$, as bands with higher $k_y$ do not couple to to the lowest nanowire band. (Note that our $y$ direction is the $x$ direction in Ref.\ \cite{Reeg2018}.)}
\end{figure}

The result of such a fitting procedure is shown in Fig.\ \ref{fig:app:dispersion_fitted_to_reeg}, where the orange segments show the dispersion obtained from our continuum model and the blue segments show the model of Ref. \cite{Reeg2018}. We conclude that the dispersions fit well. The band shift $\varepsilon_{\rm shift}$ is defined by comparing the bottom $\varepsilon_{{\rm N}0}$ of the N band to the bottom of the (quadratic) N band in the absence of the coupling to the superconductor,
\begin{equation}
\varepsilon_\textrm{shift} = \varepsilon_{\textrm{N}0} - \varepsilon_{\textrm{N}0}|_{w=\infty}.
\end{equation}
Since both models yields the same dispersion, they obviously also give the same band shift, as is visible in the lower panel of Fig.\ \ref{fig:app:dispersion_fitted_to_reeg}.

Being able to relate the results of our model to those of Ref.\ \onlinecite{Reeg2018} we now study the dependence of the band shift on the thickness $D_{\rm N}$ of the semiconductor wire and compare the band shift to the zero-point shift 
\begin{equation}
  \varepsilon_0= \frac{\hbar^2 \pi^2}{2 m_{\rm N} D_{\rm N}^2}
\end{equation}
that applies when the semiconductor wire is not coupled to the superconductor.

For weak (but nonzero) coupling between N and S the band-shift can be calculated perturbatively in $1/w$. We solve \eqref{eq:quantization_with_w} perturbatively in the single-mode regime where $\pi \lesssim \kFN D_\textrm{N} \ll 2 \pi$. We choose the expansion point $\kNz = \pi/D_\textrm{N} + \delta k$, as for large $w$, the shift $\delta k$ vanishes. To lowest order in $\delta k D_\textrm{N}$ and for  $\frac{D_\textrm{S}m_\textrm{S}}{\hbar^2\kFS}|\varepsilon_{\textrm{shift}}| \ll 1$, we find the solution
\begin{equation}
\delta k = - \frac{\hbar \pi/D_N^2 \mN}{2 w + \vFS \cot \kFS D_\textrm{S}},
\end{equation}
and the band shift
\begin{equation}\label{eq:app:band_shift}
	\varepsilon_\textrm{shift} = - \frac{\hbar^3 \pi^2/\mN^2 D_\textrm{N}^3}{ 2 w + v_\textrm{S} \cot \kFS D_\textrm{S}},
\end{equation}
where  $\vFS = \hbar \kFS/\mS$ and $\vFN = \hbar \kFN/\mN$. The band shift \eqref{eq:app:band_shift} is negative and strongly decreases as a function of $D_N$.

An intuitive understanding of the band shift can be obtained as follows. For $w \rightarrow \infty$ the transverse wavenumber is exactly quantized to $\kNz = \pi/D_N$. The confinement in the $z$ direction gives the zero-point contribution $\varepsilon_0$ to the energy $\varepsilon$. If the barrier height $w$ is decreased to a finite value, the quantization of $\kNz$ is softened and the confinement to N is gradually lifted. As a result, the zero-point contribution to the energy is decreased, which shifts the band bottom downwards in energy as compared to the case $w \to \infty$ of an intransparent interface, consistent with Eq. \eqref{eq:app:band_shift}. The band shift is bounded from below by $-\varepsilon_0$, the bound being attained when the confinement to N is fully lifted.

\begin{figure}
\includegraphics[width=\columnwidth]{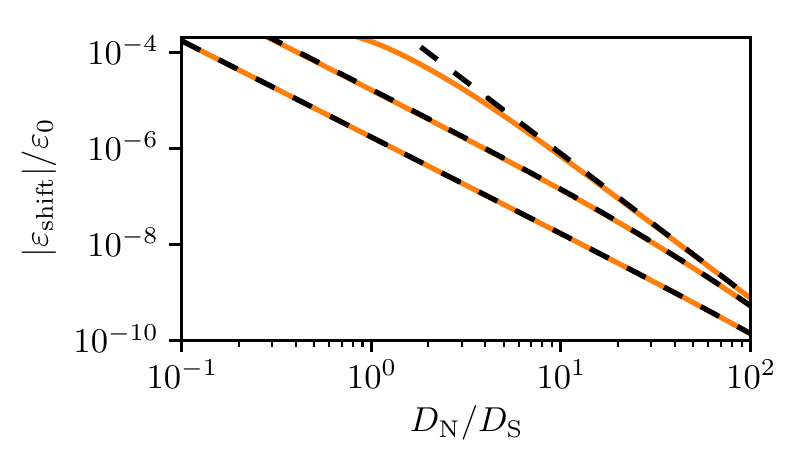}
\caption{\label{fig:app:band_shift__vs__thickness} Log-log plot of the nanowire band shift as a function of ratio $D_{\rm N}/D_{\rm S}$. The predictions from Eq. \eqref{eq:app:band_shift} (dashed lines) fit well to a numerical solution of Eq. \eqref{eq:quantization_with_w} (solid lines). The band shift decreases with $D_N$. The parameters are the same as in Fig. \ref{fig:app:dispersion_fitted_to_reeg}, except for the value of $\kFN$, for which we take $\kFN D_\textrm{N}/\pi=1$, and the height of the potential barrier at the interface, which is $w \mN/\kFN = 100$ (left durve), $w \mN/\kFN=10$ (middle) and $w=0$ (right).}
\end{figure}

\section{Numerical calculation of the scattering matrix}\label{sec:appendix:numerical_method}

We here provide details on our numerical calculation of the scattering matrix $S(L, \varepsilon)$ of a disordered superconducting wire of length $L$, attached to ideal source and drain leads. The wire is described by the Hamiltonian \eqref{eq:hamiltonian_full}; the leads are described by the same Hamiltonian, but without the disorder potential $U(\vr)$ and the superconducting order parameter $\Delta$.

To calculate $S(L, \varepsilon)$, we divide the wire for $0 < x < L$ into thin slices of length $\delta L$. We calculate the scattering matrix of a single slice using the Born approximation. The scattering matrix $S(L,\varepsilon)$ of the fully system is then obtained by concatenation of scattering matrices of the individual slices.

In order to be apply the Born approximation for the calculation of the scattering matrix of a single slice, we treat the energy $\varepsilon$, the superconducting order parameter $\Delta$, and the disorder potential $U(\vr)$ as perturbations, {\em i.e.}, we write the Bogoliubov-de Gennes Hamitonian as
\begin{equation}
  \hat {\cal H} = \hat {\cal H}_0 + \hat {\cal H}_{\varepsilon}',
\end{equation}
with
\begin{equation}
  \hat {\cal H}_0 = [\xi_{\mathbf p} + V_{\rm conf}] \tau_z,\ \
  \hat {\cal H}_{\varepsilon}' = \Delta \tau_x + U(\vr) \tau_z - \varepsilon,
\end{equation}
where the $\tau_\alpha$, $\alpha=x,y,z$ are the Pauli matrices in particle-hole space, $\xi_{\mathbf p}$ and $V_{\rm conf}$ are the kinetic energy and the confinement potential, see Sec.\ \ref{sec:model}. Application of the Born approximation is possible if $\delta L$ is small enough, $\kFS \delta L \ll k_x \xi$, $k_x \ell$, and gives
\begin{equation}
  {\cal S}_{\delta L} = [\openone - i {\cal T}_{\delta L}/2][\openone + i {\cal T}_{\delta L}/2]^{-1},\label{eq:numerical Sn}
\end{equation}
where
\begin{equation}
  ({\cal T}_{\delta L})_{\nu'\nu} =
  \int_{\delta L} d\vr
  \bra{\psi_{\nu'}(\vr_j, 0)} \hat {\cal H}'_{\varepsilon}\ket{\psi_{\nu}(\vr, 0)},\label{eq:Tmatrix definition}
\end{equation}
the integration taking place over the width of the slice. The mode functions $\psi_{\nu}(\vr,\varepsilon)$ are evaluated at zero energy, since the energy $\varepsilon$ is accounted for in the perturbation $\hat {\cal H}_{\varepsilon}'$. The multi-index $\nu = (s, \tau, n_y, \kSz)$, with $s=\pm$ and $\ket{\psi_\nu}$ is defined in Eq. \eqref{eq:metal_metal_wave_function}. Equation \eqref{eq:Tmatrix definition} implements the first order Born approximation, while at the same time preserving unitarity of the scattering matrix ${\cal S}_{\delta L}$. 

We conclude by presenting explicit expressions for the $T$-matrix of Eq.\ \eqref{eq:Tmatrix definition}. We separate the $T$ matrix into three contributions,
\begin{equation}
	{\cal T}^{(j)}_{\nu'\nu} = {\cal T}^{(j)}_{\varepsilon, \nu'\nu} + {\cal T}^{(j)}_{\Delta, \nu'\nu} + {\cal T}^{(j)}_{\gamma,\nu'\nu}.
\end{equation}
The first two contributions read
\begin{align}
	{\cal T}^{(j)}_{\varepsilon, \nu'\nu} =& 
		-\varepsilon \delta_{\nu', \nu} \chi_{\nu', \nu}
	,\\
	{\cal T}^{(j)}_{\Delta, \nu'\nu} =&
		\Delta
		\chi_{\nu', \nu} \Theta_{\nu', \nu},
\end{align}
where
\begin{align}
	\chi_{\nu', \nu}  =& -i 
	    \frac{
			e^{i q_{\nu', \nu} x}
			\left[ e^{i q_{\nu', \nu} \delta L} -1 \right]  	    
	    }{
	    	\hbar q_{\nu', \nu}\sqrt{v_x'v_x} 
	    },\\
	    q_{\nu', \nu} =& \tau s k_x- \tau' s' k_x',\\
	\Theta_{\nu', \nu} =&
		\frac{
			2 D_{\rm S} \tau' \tau d_{\tau'}^* d_{\tau} 
			e^{i (\tau' \kSz' - \tau \kSz) D_{\rm S}}
		}{
			\sqrt{\mathcal{N}_{\nu'} \mathcal{N}_\nu \vSz' \vSz}
		}
		\times
		\\&
		\nonumber
		\left[\sinc D_{\ell} (\kSz' - \kSz) - \sinc D_{\ell} (\kSz' + \kSz)\right],
\end{align}
with $k_x = k_x(0)$. The third contribution, which describes scattering from the disorder potential $U(\vr)$, takes the form
\begin{align}
	{\cal T}^{(j)}_{\gamma, \nu'\nu} =& 4 \tau  \sqrt{\gamma}
	 \delta_{\tau', \tau} 
		e^{i q_{\nu', \nu} x_i}
		(X^{(i)}_{\nu', \nu} + i Y^{(i)}_{\nu', \nu}) \times
		\\
		&\frac{
			d_{\tau'}^* d_\tau e^{- i \tau (\kSz - \kSz') D_{\textrm{S}}}
		}{
			\sqrt{\mathcal{N}_{\nu'} \mathcal{N}_{\nu} \vSz \vSz' v_x v_x'}
		}.
\end{align}
Here $\gamma = \hbar v_{\rm S}/2 \pi \nu_0 \ell$, and $X^{(i)}_{\nu', \nu}$ and $Y^{(i)}_{\nu', \nu}$ are correlated Gaussian random variables with zero mean. The covariance matrix of $(X_{\nu', \nu}, Y_{\nu', \nu})^{\rm T}$ reads
\begin{equation}\label{eq:cov_mat}
C = 
\begin{pmatrix}
C^{(x)}_{XX} & C^{(x)}_{XY} \\
(C^{(x)}_{XY})^{\rm T} & C^{(x)}_{YY}
\end{pmatrix}
 C^{(y)} C^{(z)},
\end{equation}
where we dropped the indices $(\nu_1', \nu_1), (\nu_2', \nu_2)$ that are attached to each $C$, using the convention that the different $C$s are multiplied element wise. For three dimensions, the explicit forms of the coefficients in the covariance matrix is
\begin{widetext}
\begin{align}
C^{(x)}_{XX, (\nu'_1, \nu_1), (\nu'_2, \nu_2)} &= 
	\frac{\delta L}{2} \left[\sinc \delta L (q_1 - q_2) + \sinc \delta L (q_1 + q_2) \right],
\\
C^{(x)}_{XY, (\nu'_1, \nu_1), (\nu'_2, \nu_2)} &= 
	\frac{q_2 - q_1}{4} \delta L^2 \left[\sinc \delta L (q_1 - q_2)/2 \right]^2
	+
	\frac{q_2 + q_1}{4} \delta L^2 \left[\sinc \delta L (q_1 + q_2)/2 \right]^2,
\\
C^{(x)}_{YY, (\nu'_1, \nu_1), (\nu'_2, \nu_2)} &= 
	\frac{\delta L}{2} \left[ 
		\sinc \delta L (q_1 - q_2) - \sinc \delta L (q_1 + q_2)
	\right],
\\ 
C^{(y)}_{(\nu'_1, \nu_1), (\nu'_2, \nu_2)} &= 
	\frac{1}{2 D_{\rm N}} 
	\left[
		\delta_{0, n_{y, 1}' + n_{y, 1} - n_{y, 2}' - n_{y, 2}} 
		+
		\delta_{0, n_{y, 1}' - n_{y, 1} - n_{y, 2}' + n_{y, 2}} 
		+
		\delta_{0, n_{y, 1}' - n_{y, 1} + n_{y, 2}' - n_{y, 2}} 
	\right.		
		\nonumber\\
		&
	\left.
		-
		\delta_{0, n_{y, 1}' - n_{y, 1} - n_{y, 2}' - n_{y, 2}} 
		-
		\delta_{0, n_{y, 1}' - n_{y, 1} + n_{y, 2}' + n_{y, 2}} 
		-
		\delta_{0, n_{y, 1}' + n_{y, 1} - n_{y, 2}' + n_{y, 2}} 
	\right.		
		\nonumber\\
		&
	\left.
		-
		\delta_{0, n_{y, 1}' + n_{y, 1} + n_{y, 2}' - n_{y, 2}} 
	\right],
\\
C^{(z)}_{(\nu'_1, \nu_1), (\nu'_2, \nu_2)} &= 
	\frac{D_{\rm S}}{8}
	\left[
		\sinc D_{\ell}(k_{{\rm S}z1}' + k_{{\rm S}z1} + k_{{\rm S}z2}' + k_{{\rm S}z2})
		+
		\sinc D_{\ell}(k_{{\rm S}z1}' + k_{{\rm S}z1} - k_{{\rm S}z2}' - k_{{\rm S}z2})
	\right.
	\nonumber\\
	&\left.
		+
		\sinc D_{\ell}(k_{{\rm S}z1}' - k_{{\rm S}z1} - k_{{\rm S}z2}' + k_{{\rm S}z2})
		+
		\sinc D_{\ell}(k_{{\rm S}z1}' - k_{{\rm S}z1} + k_{{\rm S}z2}' - k_{{\rm S}z2})
	\right.
	\nonumber\\
	&\left.
		-
		\sinc D_{\ell}(k_{{\rm S}z1}' - k_{{\rm S}z1} - k_{{\rm S}z2}' - k_{{\rm S}z2})
		-
		\sinc D_{\ell}(k_{{\rm S}z1}' - k_{{\rm S}z1} + k_{{\rm S}z2}' + k_{{\rm S}z2})
	\right.
	\nonumber\\
	&\left.
		-
		\sinc D_{\ell}(k_{{\rm S}z1}' + k_{{\rm S}z1} - k_{{\rm S}z2}' + k_{{\rm S}z2})
		-
		\sinc D_{\ell}(k_{{\rm S}z1}' + k_{{\rm S}z1} + k_{{\rm S}z2}' - k_{{\rm S}z2})
	\right],
\end{align}
\end{widetext}
where $q_i = q_{\nu_i', \nu_i}$, $\delta_{n, m}$ is the Kronecker delta and $\sinc x = (\sin x)/x$. For two dimensions the $C^{(x)}$ and $C^{(y)}$ are the same as in three dimension. For the $y$ direction we restrict the mode indices to $n_y = 1$ and we set $C^{(y)}$ equal to the identity matrix.

\FloatBarrier\bibliographystyle{apsrev4-1}

\end{document}